**LAYER: A Quantitative Explainable AI Framework for Decoding Tissue-Layer Drivers of Myofascial Low Back Pain**

Zixue Zeng [1, 2], Anthony M. Perti[1], Tong Yu[1, 2], Grant Kokenberger[1], Hao-En Lu[1], Jing Wang[1], Xin Meng[1], Zhiyu Sheng[3,8], Maryam Satarpour[1], John M. Cormack[3, 8], Allison C. Bean[7,9,10], Ryan P. Nussbaum[7], Emily Landis-Walkenhorst[4], Kang Kim[1, 3, 8, ‡], Ajay D. Wasan[4, 5, ‡], Jiantao Pu[1, 2, 6, *, ‡]

[1]Department of Bioengineering, University of Pittsburgh, Pittsburgh, PA 15213, USA

[2]Department of Radiology, University of Pittsburgh, Pittsburgh, PA 15213, USA

[3]Cardiology, Department of Medicine, University of Pittsburgh, PA 15213, USA

[4]Department of Anesthesiology & Perioperative Medicine, University of Pittsburgh, School of Medicine, PA 15213, USA

[5]Department of Psychiatry, University of Pittsburgh, School of Medicine, PA 15213, USA

[6] Department of Ophthalmology, University of Pittsburgh, Pittsburgh, PA 15213, USA

[7] Department of Physiatry, University of Pittsburgh School of Medicine, PA 15213, USA

[8]Vascular Medicine Institute, University of Pittsburgh, PA 15213, USA

[9]Bethel Family Musculoskeletal Research Center, University of Pittsburgh, PA 15213, USA

[10]McGowan Institute for Regenerative Medicine, PA, 15219, USA

**\* Corresponding authors and guarantors of the entire manuscript.**

**‡ These authors are equal senior co-authors and co–principal investigators of the project.**

Jiantao Pu, PhD
3240 Craft Place
Pittsburgh, PA 15213
jip13@pitt.edu
(412) 641-2571

**Abstract**

Myofascial pain (MP) is a leading cause of chronic low back pain, yet its tissue-level drivers remain poorly defined and lack reliable image biomarkers. Existing studies focus predominantly on muscle while neglecting fascia, fat, and other soft tissues that play integral biomechanical roles. We developed an anatomically grounded explainable artificial intelligence (AI) framework, LAYER (Layer-wise Analysis for Yielding Explainable Relevance Tissue), that analyses six tissue layers in three-dimensional (3D) ultrasound and quantifies their contribution to MP prediction. By utilizing the largest multi-model 3D ultrasound cohort consisting of over 4,000 scans, LAYER reveals that non-muscle tissues contribute substantially to pain prediction. In B-mode imaging, the deep fascial membrane (DFM) showed the highest saliency (0.420), while in combined B-mode and shear-wave images, the collective saliency of non-muscle layers (0.316) nearly matches that of muscle (0.317), challenging the conventional muscle-centric paradigm in MP research and potentially affecting the therapy methods. LAYER establishes a quantitative, interpretable framework for linking layer-specific anatomy to pain physiology, uncovering new tissue targets and providing a generalizable approach for explainable analysis of soft-tissue imaging.

## 1. Introduction

Low back pain is the second most common medical condition worldwide, surpassed only by the common cold [1]. An estimated 70-85% of people will experience low back pain at some point in their lives [2], with 5-15% of cases progressing to become chronic lower back pain (cLBP)[3]. Despite extensive research, there is still no standardized, reliable, and quantitative method to identify the biomechanical factors driving cLBP [4].

Myofascial pain (MP) is a major contributor to cLBP, arising from abnormalities across multiple soft tissue layers and the formation of tender points and "trigger points" within taut bands of skeletal muscle and surrounding fascia [5,6]. Previous studies have investigated individual tissue layers in isolation, such as increased stiffness in lumbar muscles [7], reduced activation in the transversus abdominis (TrA) [8], or thickened and less mobile fascial membranes [9]. Few have analyzed their combined or interactive roles. This fragmented approach overlooks potential contributions from adjacent tissues, including superficial and deep fat layers, and their mechanical coupling with fascia and muscle. Moreover, most analyses rely on limited handcrafted features, such as tissue stiffness or thickness, which cannot fully capture the spatial and mechanical complexity underlying MP. Thus, a comprehensive, layer-specific quantitative analysis incorporating all relevant soft-tissue layers could yield a more accurate and mechanistic understanding of MP and enable target, tissue-specific interventions [10,11].

Most existing tissue layer analysis studies rely on two-dimensional (2D) ultrasound to assess tissue thickness, contraction, and echogenicity associated with MP [12]. While informative, 2D imaging lacks volumetric context and may overlook out-of-plane abnormalities that are common in multifocal pain regions. In contrast, three-dimensional (3D) ultrasound provides richer spatial information, enabling visualization of tissue continuity and deformation patterns across depth. However, interpreting 3D ultrasound is technically challenging due to limited resolution, speckle noise, and the inherently fuzzy boundaries between adjacent layers, making manual delineation subjective and labor-intensive [13]. Artificial intelligence (AI) offers the ability to automatically extract high-dimensional, spatially contextual features from complex ultrasound data, yet the AI models often act as "black boxes" with limited interpretability. Explainable AI (XAI) methods such as gradient-based saliency map can visualize regions that influence predictions, but these approaches remain largely qualitative, fail to quantify the relative contribution of anatomical layers, and are sensitive to image noise and acquisition variability [14-20]. As a result, prior approaches cannot reliably link AI-derived features to biologically or clinically meaningful tissue structures.

In this study, we introduce a novel, anatomically grounded XAI framework, termed Layer-wise Analysis for Yielding Explainable Relevance Tissue (LAYER), to quantitatively determine how individual tissue layers contribute to MP. LAYER employs a region-wise occlusion strategy, comparing model logits before and after selectively masking specific tissue layers to derive quantitative saliency scores. This approach moves beyond qualitative heatmaps to provide objective, reproducible measures of anatomical relevance. A multi-layer occlusion extension further reveals whether tissue pairs exhibit synergistic or redundant effects. To improve feature learning, we designed a curriculum-adaptive re-weighting network (CARN) that combines self-paced curriculum learning with adaptive intensity re-weighting (AIR). In CARN, a 3D U-Net generates per-voxel weight maps that guide a 3D ResNet to classify, enabling adaptive focus on anatomically informative regions during training.

We analyzed a prospective cohort of 113 participants (72 with cLBP, 41 healthy controls) from the **M**ultimodal **U**ltrasound-based **S**ignature and **C**omputational **L**earning for **E**valuation of Myofascial Pain Study (MUSCLE study). It contains over 4,000 co-registered multimodal (B-mode and shear-wave) 3D ultrasound scans, which, to our knowledge, is the largest dataset of its kind. LAYER identified that the deep fascial membrane (DFM) was most informative in B-mode, whereas muscle layers dominated in shear-wave imaging. When modalities were fused, non-muscle layers collectively contributed comparably to muscle, underscoring the interdependence of fascia, fat, and muscle in MP. LAYER outperformed common XAI baselines (e.g., Grad-CAM) in faithfulness metrics and produced tissue-specific, anatomically consistent explanations that aligned with known biomechanical relationships. Overall, this study introduces the first quantitative, multi-layer, and multimodal XAI framework for analyzing 3D ultrasound in MP. By linking AI predictions directly to anatomically segmented tissue layers, LAYER provides interpretable, biologically meaningful, and clinically actionable insights into the mechanisms of chronic low back pain, laying the foundation for precision strategies and development of tissue-targeted rehabilitation treatments.

## 2. Results

### 2.1 Cohort design, 3D ultrasound acquisition, and annotation

We established what is, to our knowledge, the largest 3D ultrasound cohort for studying MP in the lower back. This dataset comprises 113 participants (72 with cLBP and 41 healthy) and 2,613 B-mode scans, of which 1,392 have co-registered shear-wave elastography (SWE) acquisitions. This study aimed to characterize tissue-level correlates of lumbar MP using 3D ultrasound imaging.

Participants were positioned prone, and volumetric scans were acquired using an ultrasound array transducer placed laterally to the midline at the L3–L4 interspace, targeting the multifidus (MF) and erector spinae (ES) muscles (Fig. 1 A). A synthetic aperture approach with temporal compounding was used to capture 3D volumes. Imaging was performed bilaterally, with two sites identified per side over the MF and ES by board-certified physiatrists or pain specialists experienced in musculoskeletal ultrasound. Each site was imaged three times per clinical visit, yielding a total of 12 B-mode volumes per participant per visit. Co-registered SWE scans were obtained at the same locations with two repetitions per site.

To diagnose MP and identify tender and trigger points, participants were examined in the prone position while a physician palpated from L5 to L3 to locate painful myotomes and taut bands extending laterally to the mid-flank. Trigger points were identified as pain locations with radiating pain, since in the lumbar spine, it is often not possible to identify taut bands or nodules. Tender points were identified as localized pain without referred pain. The pain pressure threshold (PPT) and patient-reported pain level (0-10) were recorded for quantitative confirmation. Compared with control sites, MP regions showed lower PPT values ($4.08 \pm 0.27$ kg/cm² vs. $6.66 \pm 0.35$ kg/cm², $p < 0.001$) and higher pain scores ($5.35 \pm 0.31$ vs. $3.60 \pm 0.28$, $p < 0.001$) (Fig. 1 D). Ultrasound scans were labeled as healthy control if no tender or trigger points were detected, or as MP-positive if at least one tender or trigger point was identified. MP-positive scans were further sub-categorized into tender MP ($\geq 1$ tender point, no trigger points) or trigger-point MP ($\geq 1$ trigger point, with or without tender points). At the visit level, a participant was considered MP-positive if any tender or trigger points were detected bilaterally; otherwise, the visit was classified as control.

To assess generalizability and simulate real-world deployment, the cohort was split at the patient level into six folds: five folds for model development and cross-validation, and one independent fold held out for final testing. All scans from the same participant, including both sides and all visits, were confined to a single fold to prevent data leakage.

## 2.2 LAYER workflow

Each participant's 3D ultrasound was analyzed using LAYER, a three-phase computational pipeline (Fig. 1 E) designed to generate anatomically interpretable predictions from multimodal ultrasound images:

**(1) Image Segmentation phase:** A Generative Reinforcement Network (GRN) automatically segmented six anatomically distinct tissue layers from each B-mode volume, including dermis, superficial fascial layer (SFL), super fascial membrane (SFM), deep fat, DFM, and muscle [21].

**(2) CARN training phase:** CARN served as the ultrasound scan classifier, integrating self-paced curriculum learning and AIR to stabilize training and emphasize anatomically informative regions.

**(3) Interpretation phase:** Model interpretability was achieved through a layer-wise occlusion analysis that quantified the contribution of each tissue layer to the prediction. For each scan, we measured the change in model logit output after occluding a single layer, yielding a tissue-layer saliency score that reflects its relative importance (Fig. 2 A). A volume-adjusted saliency score was computed to normalize for layer size and suppress background noise.

To interpret directionality, we calculated the Positive Directional Saliency Score (PDSS), indicating how often a layer's presence increased the probability of MP, and the Negative Directional Saliency Score (NDSS), indicating the opposite effect toward control predictions. For the inter-layer relationship, we performed a multi-layer occlusion analysis to compute an occlusion-interaction score (OIS) (Fig. 2 C), where positive values (OIS > 0) denote synergy—the joint occlusion effect exceeding the sum of individual effects—and negative values (OIS < 0) denote redundancy between tissue layers.

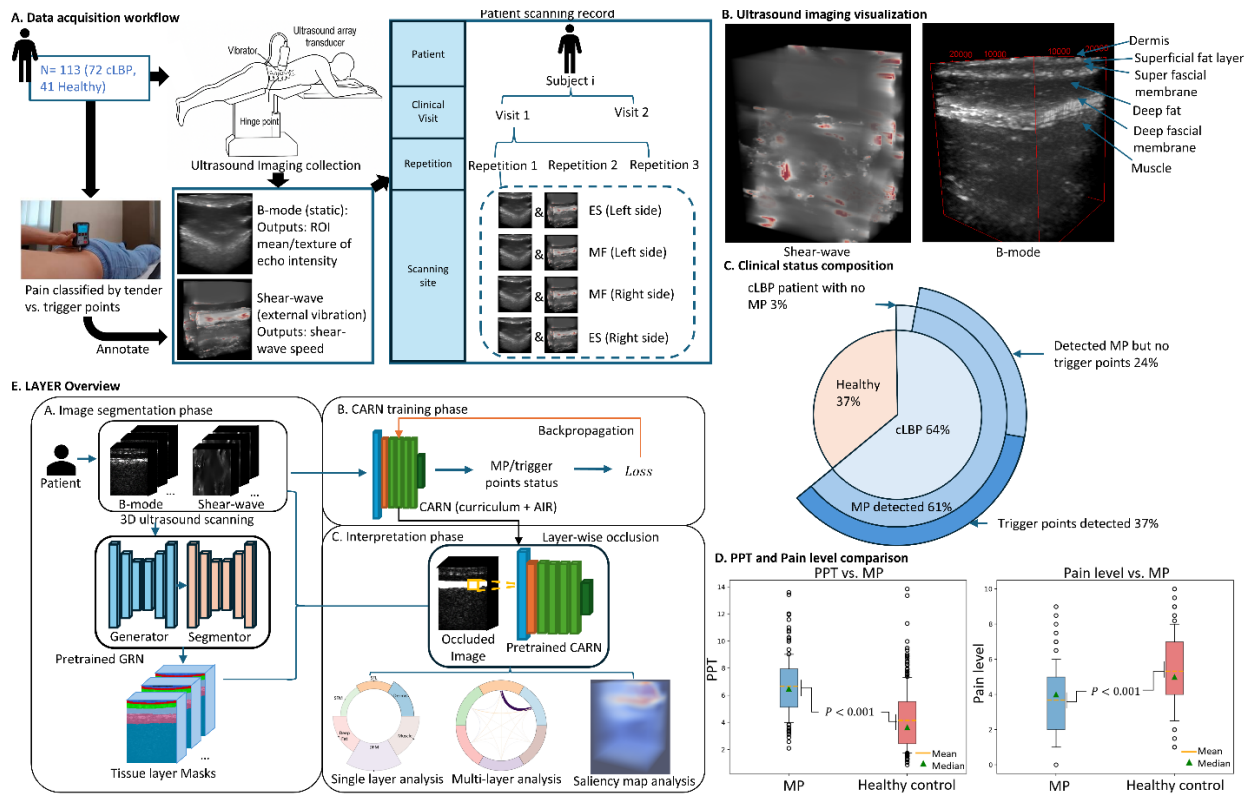

**Figure 1. Overview of cohort design, imaging, and LAYER workflow. A.** Data acquisition and labeling. Participants underwent 3D B-mode and co-registered SWE imaging. Clinicians identified and labeled tender and trigger points through standardized palpation. Each participant completed two clinical visits. During each visit, four sites over the MF and ES muscles were scanned three times, yielding 12 B-mode volumes per visit. **B.** Representative B-mode and SWE images with the six automatically segmented tissue layers: dermis, superficial fat, SFM, deep fat, DFM, and muscle. **C.** Cohort composition showing the distribution of clinical status: 61% of participants presented with MP, and 37% had identifiable trigger points. **D.** Clinical validation metrics. Boxplots comparing PPT and self-reported pain scores between MP and control regions. MP regions exhibited significantly lower PPT and higher pain levels (p < 0.001), confirming the reliability of clinically identified pain points. **E.** LAYER computational pipeline. The pretrained GRN first segments six tissue layers from B-mode images. The CARN is then trained on multimodal ultrasound volumes to classify MP and trigger points. Finally, a layer-wise occlusion analysis quantifies each tissue layer's contribution to model predictions, generating anatomically grounded saliency maps.

## 2.3 Layer-specific contributions differ by imaging modality

Layer-wise occlusion analysis reveals distinct saliency score distribution between B-mode and SWE (Fig. 2 B). In B-mode, the DFM exhibited the highest raw saliency score (0.420, 95% confidence interval (CI) 0.413–0.427), followed by muscle (0.279) and deep fat (0.209) (Supplementary Table 2). After normalizing for tissue volume, superficial layers became dominant: dermis ($37.83 \times 10^{-7}$, 36.44–39.22) and superficial fat ($21.15 \times 10^{-7}$, 20.22–22.08) showed the highest volume-adjusted saliency score, indicating that predictive information in B-mode is concentrated within thin superficial tissues. Although the muscle displayed high raw saliency, its normalized score was lowest, suggesting that its apparent importance primarily reflects its large volume rather than distinctive image features. In SWE, the attention pattern shifted toward deep tissues. The muscle layer showed the highest saliency score by a wide margin (0.625, 0.596–0.654), followed by DFM (0.180, 0.171–0.189). Volume-adjusted saliency scores preserved this ordering, with muscle again leading ($5.21 \times 10^{-7}$, 4.97–5.45), confirming that SWE captures richer mechanical information from deep structures. When B-mode and SWE were combined, muscle retained the largest single-layer contribution (0.317, 0.287–0.347), yet the aggregate of all non-muscle layers was comparable (sum = 0.316), highlighting the distributed and complementary contributions of superficial and deep tissues.

Directional saliency metrics further supported these modality-dependent effects (Supplementary Table 3). In B-mode, DFM strongly pushed predictions toward MP (PDSS = 0.271), whereas in SWE, muscle dominated the MP-driving signal (PDSS = 0.639), and DFM instead tended to favor control predictions (NDSS = 0.089). Volume-adjusted PDSS reinforced this contrast: B-mode derives disproportionate predictive signal from superficial layers (e.g., dermis $10.56 \times 10^{-7}$), while the SWE concentrated discriminative information within deep tissues (muscle $5.33 \times 10^{-7}$).

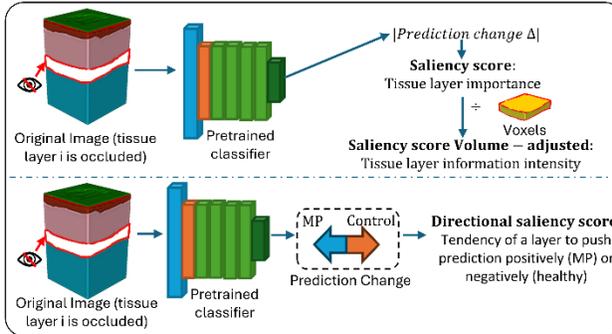

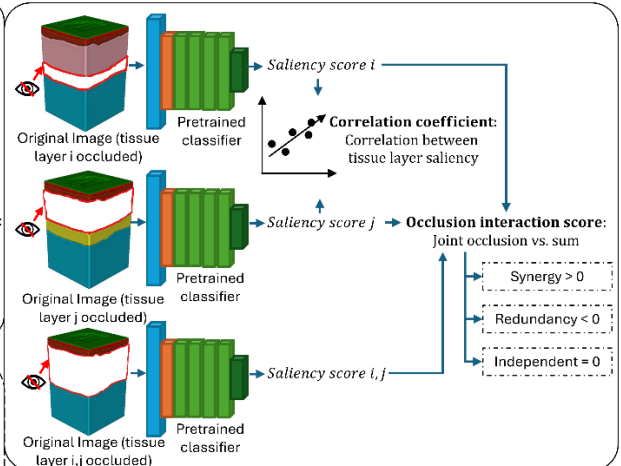

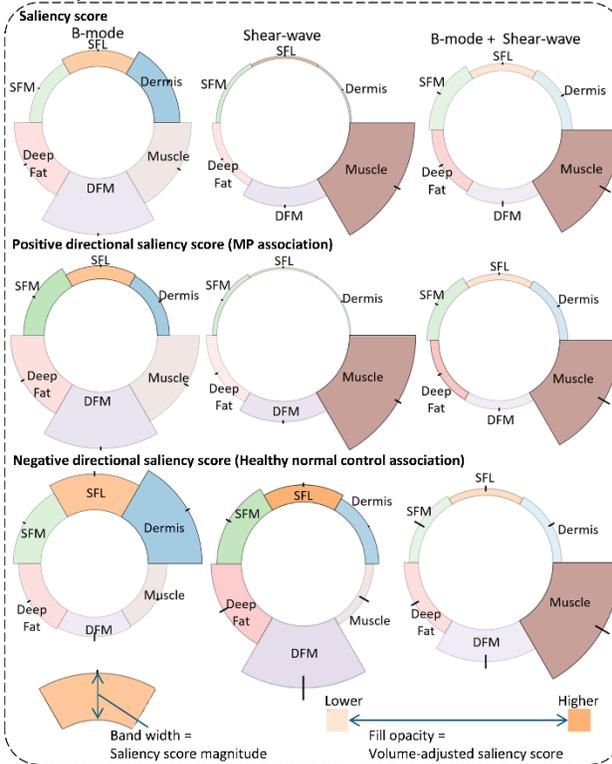

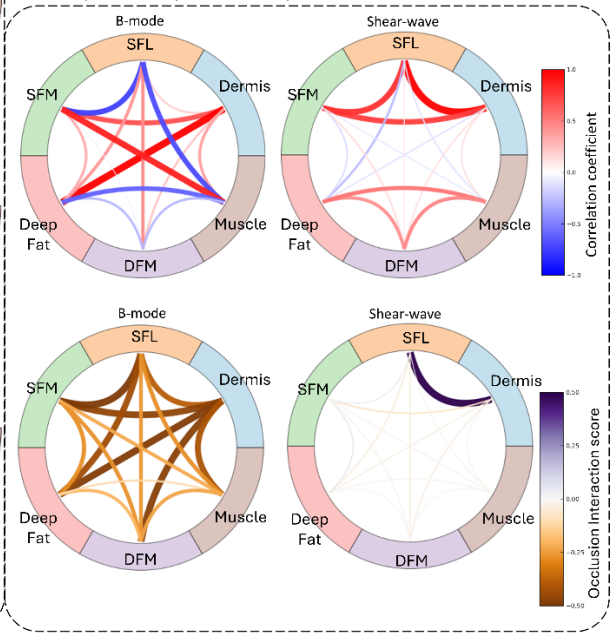

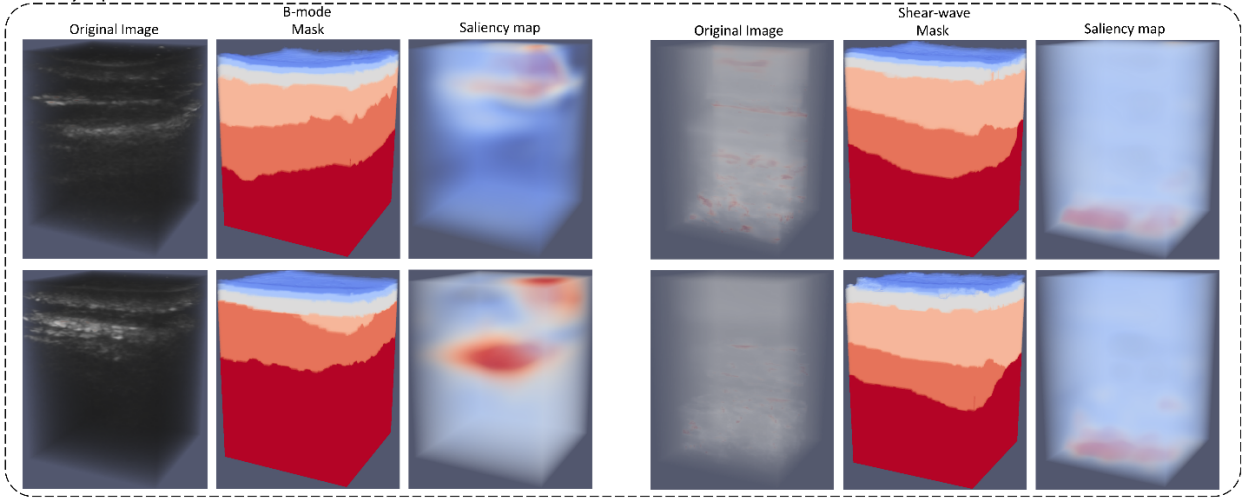

**Figure 2. Quantitative layer-wise and inter-layer explainability analysis using LAYER. A.** Single-layer occlusion analysis. For each scan, one tissue layer was occluded and reevaluated by the model. The resulting change in prediction defined the layer's saliency score (layer importance). Scores were normalized by tissue layer volume to yield volume-adjusted saliency (information intensity). Directional saliency score was also computed to indicate whether a layer's presence increased the likelihood of MP or control classification. **B.** Saliency annulus visualization. The circular diagram is divided into six segments, representing dermis, SFL, SFM, Deep fat, DFM, and muscle. Band thickness reflects the mean saliency scores, with radial black ticks indicating the 95% CI. Fill opacity encodes volume-adjusted saliency scores (darker = higher). In B-mode images, DFM exhibited the greatest saliency, whereas in SWE, the muscle layer dominates. For the multimodal image of B-mode and SWE, the cumulative contribution of non-muscle layers is comparable to that of muscle, indicating their collective and non-negligible role in MP. **C.** Multi-layer occlusion analysis. Pairs of tissue layers were simultaneously occluded to evaluate inter-layer dependencies. Correlations between layer-wise saliency scores and an OIS were calculated to determine whether joint occlusion produced effects greater (synergy, OIS > 0) or smaller (redundancy, OIS < 0) than the sum of individual occlusions. **D.** Chorded saliency annulus of inter-layer relationships. The outer ring corresponds to the six tissue layers as in panel B. Interior chords represent either saliency correlations or OIS values, with thicker chords denoting stronger relationships. In SWE, two prominent clusters, including SFL/dermis/SFM and muscle/DFM/deep fat, were observed, each exhibiting strong intra-cluster correlation. **E.** Representative ultrasound scans and saliency maps overlays. Saliency is concentrated within muscle regions in SWE, whereas fascial layers exhibit stronger contributions in B-mode.

## 2.4 Multi-layer interactions and spatial coupling

Multi-layer analyses revealed interdependence among tissue-layer saliency scores and their spatial organization (Fig. 2 D). The chorded saliency annulus showed two coherent clusters in SWE: a superficial cluster (dermis + SFL + SFM) and a deep cluster (deep fat + DFM + muscle). This organization suggests that anatomically adjacent tissue layers act as mechanically coupled units, exhibiting coordinated relevance patterns in model predictions. In contrast, B-mode imaging showed weaker correlations among deep tissues, reflecting its greater sensitivity to superficial echogenic and fascial features rather than to mechanical coupling in deeper structures. The OIS further distinguished synergy (OIS > 0) from redundancy (OIS < 0) between layer pairs. In SWE, DFM–muscle pairs exhibited positive OIS values, indicating additive contributions to MP, whereas superficial pairs (dermis–SFL and SFL–SFM) displayed negative OIS values, suggesting overlapping or redundant feature representations.

## 2.5 LAYER outperforms standard XAI with biological fidelity

We benchmarked LAYER against three widely used post-hoc explainability methods (Grad-CAM, Integrated Gradients (IG), and SmoothGrad) using insertion/deletion–based faithfulness tests performed across cross-validation folds. For each method, tissue layers were ranked by saliency scores, and two complementary perturbation procedures were applied: **(1) Insertion**, in which layers were cumulatively added to an empty volume while recording changes in the target logit; and **(2) Deletion**, in which layers were successively removed from the full volume. Methods that correctly prioritize truly influential tissues are expected to yield a steeper insertion curve, a sharper deletion drop, a higher area under the insertion curve (AUC-insertion) and an increase in Region of Focus (IROF), and a lower AUC-deletion.

Across all folds, LAYER consistently outperformed alternative XAI methods, achieving the highest AUC-insertion (1.384, 95% CI 1.289–1.478), IROF (1.055, 0.941–1.168), and AUC-$\Delta$ (0.545, 0.446–0.639), together with the lowest AUC-deletion (0.839, 0.772–0.906) (Supplementary Table 4). Paired t-tests confirmed statistically significant improvement across all four metrics ($p < 0.05$). Correspondingly, the insertion/deletion curves (Fig. 3 B) showed that revealing or removing the top-ranked layers identified by LAYER produced the largest prediction shifts, confirming the method's superior sensitivity to anatomically relevant regions.

To assess attribution dependency on learned model parameters, we performed a model-randomization sanity test by re-initializing the classifier with random weights and recomputing saliency maps. Saliency magnitudes collapsed by more than $10^3$-fold (mean saliency 0.125 vs. 0.0000845 after randomization) (Supplementary table 5), and layer-specific scores dropped substantially (Fig. 3 C), indicating that LAYER's explanations arise from learned representations rather than spurious image structure.

We next examined directional consistency between saliency and clinical status by relating the positive and negative directional saliency scores (PDSS and NDSS) to side-level MP labels. Strong associations indicate that adding a given tissue layer shifts predictions appropriately, toward MP for PDSS or toward healthy control for NDSS, demonstrating biologically meaningful feature attribution. Five of six tissue-layer directional saliency scores were significantly associated with MP ($p<0.05$), supporting the reliability of LAYER in identifying clinically relevant, layer-specific image features.

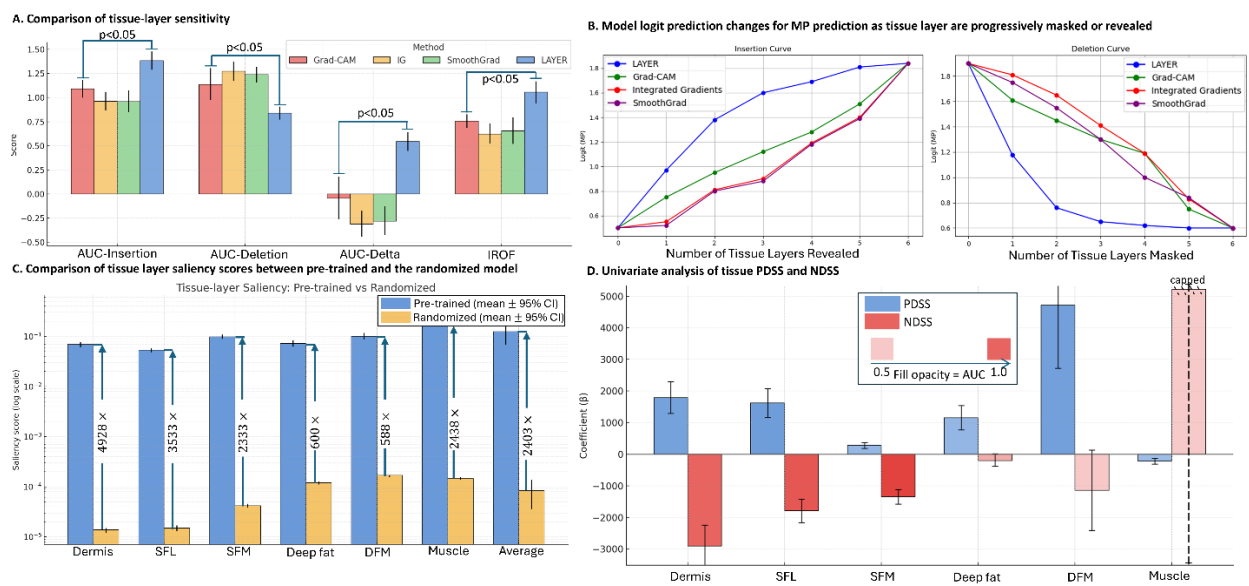

**Figure 3. Quantitative benchmarking and validation of LAYER explanations. A.** Comparison of tissue-layer sensitivity across XAI methods. LAYER achieved higher AUC-Insertion, AUC-Delta, and IROF, and lower AUC-Deletion than three widely used XAI approaches (Grad-CAM, Integrated Gradients, and SmoothGrad), indicating superior ability to rank tissue layers that drive MP predictions. **B.** Insertion and deletion curves for MP prediction. Model logits were recorded as tissue layers were progressively revealed (insertion) or masked (deletion), ordered from highest to lowest saliency according to each XAI method. LAYER produced the steepest insertion rise and sharpest deletion drop, demonstrating greater faithfulness in identifying influential tissue layers. **C.** Model-randomization sanity test. Comparison of tissue-layer saliency scores between the pre-trained classifier and a randomized model using concatenated B-mode SWE inputs. After randomization of model weights, average saliency scores decreased by more than 2400-fold, confirming that LAYER's attributions depend on learned representations rather than image structure. **D.** Clinical association of directional saliency scores. Univariate analysis relating PDSS and NDSS to side-level MP status for B-mode models. All layers except muscle exhibited significant coefficients (p<0.05), indicating that LAYER captures discriminative, layer-specific patterns consistent with clinical findings.

## 2.6 Predictive performance and scan-level aggregation

The CARN model integrates a self-paced curriculum scheduler, which prioritizes easier scans during training, with a 3D U-Net weight generator that produces voxel-wise intensity weights for a 3D ResNet classifier. During inference, scan-level probabilities from the same side or clinical visit were averaged to generate aggregated predictions, enhancing stability and clinical interpretability. Model performance for MP and trigger-point prediction was evaluated via cross-validation under three input configurations: B-mode only, SWE only, and multimodal (concatenated B-mode and SWE) (Fig. 4 B, Table 6). Using the multimodal image input, CARN achieved an average area under the curve (AUC) of 0.8210 (standard deviation (SD): 10.27) for patient-level MP prediction and 0.7804 (SD: 8.04) for side-level MP prediction (Supplementary Table 7). Compared with single-modality models, the multimodal input improved per-visit AUC by 2.93% over B-mode alone and 9.05% over SWE alone. Side-level aggregation further increased AUC by 6.58% relative to unaggregated predictions, demonstrating that temporal and spatial averaging can enhance robustness. On the independent test dataset (Fig. 4. C), the multimodal model achieved an average AUC of 0.8538 for patient-level MP classification.

Benchmarking against alternative classifier (Fig. 4 E) showed that CARN significantly outperformed other state-of-the-art models ($p < 0.05$, DeLong's test). It exceeded the 3D-CNN baseline by 1.83% in AUC ($p < 0.05$, DeLong's test) and gained an additional 4.60% improvement with curriculum learning enabled. When compared with logistic-regression baselines using only clinical features—PPT or pain rating—CARN achieved higher AUCs, surpassing the PPT model by 2.37% in cross-validation and 5.3% on the independent test set (Supplementary Table 8).

In the clinical-utility analysis, visit-level aggregation yielded an AUC of 0.853, outperforming both the PPT model (0.800) and the pain-rating model (0.776). A combined model integrating CARN + PPT + pain rating achieved an AUC of 0.995, indicating that ultrasound–derived features provide substantial complementary information and may meaningfully augment clinical assessment of MP.

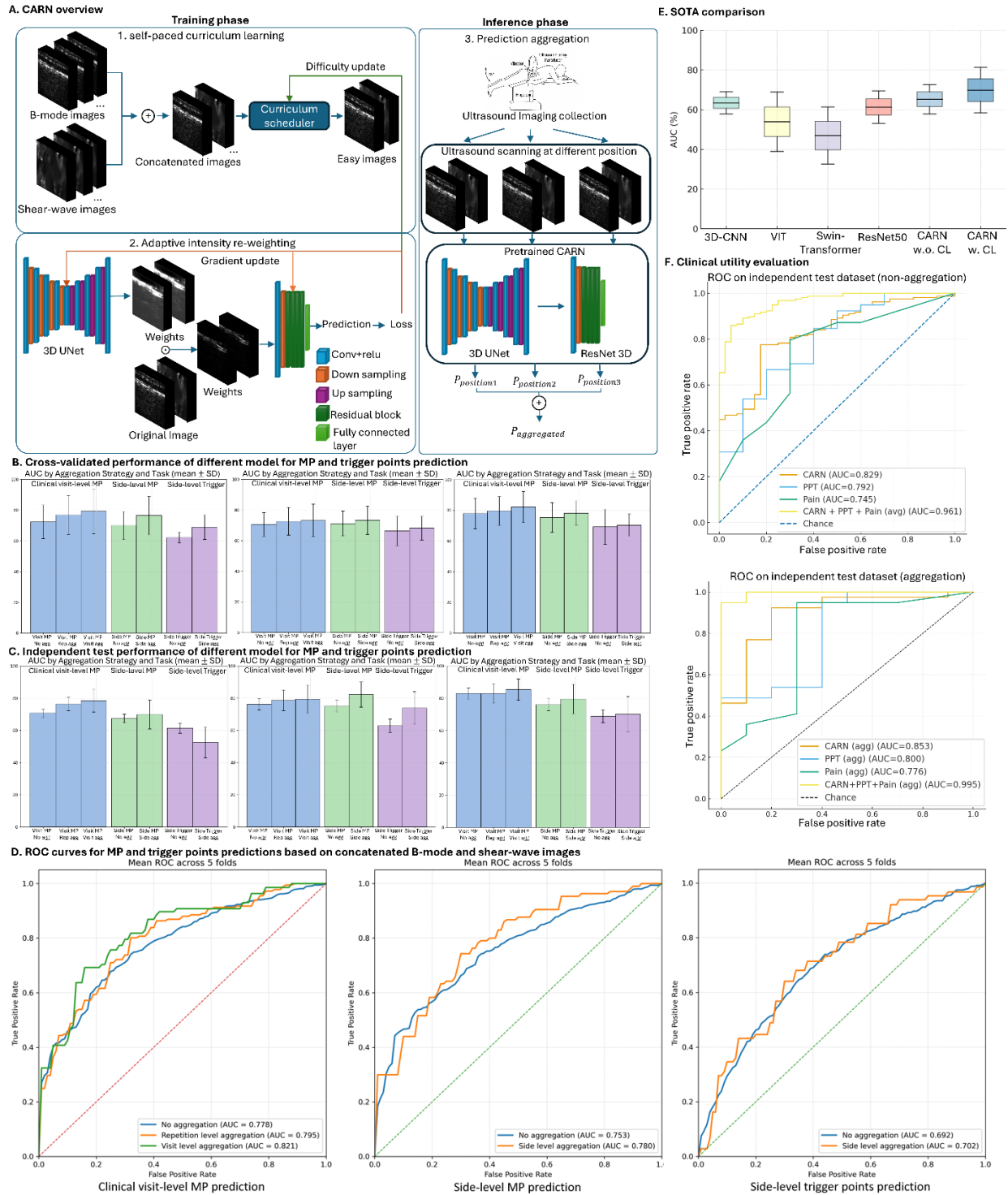

**Figure 4. Architecture, performance, and clinical utility of the CARN. A.** Network architecture. B-mode and SWE are concatenated and processed through a self-paced curriculum scheduler, which prioritizes easier samples to accelerate convergence. The concatenated input is passed to a 3D U-Net–based weight generator to produce voxel-wise intensity weights for a 3D ResNet classifier for predicting clinical status. Loss gradients are backpropagated jointly to the weight generator, classifier, and scheduler to refine both spatial weighting and sample difficulty. During inference, predictions from multiple scans of the same participants

are aggregated to produce the final probability score. **B.** Cross-validated model performance. CARN achieved an average AUC of 0.8210 (five-fold cross-validation) for MP and trigger point prediction using concatenated B-mode and SWE inputs, outperforming single-modality models. **C.** Independent test performance. On the held-out test dataset, CARN achieved an AUC of 0.854 for patient-level MP classification, confirming strong generalizability. **D.** Receiver-operating characteristic (ROC) curves. ROC curves for MP and trigger points predictions based on concatenated B-mode and SWE inputs. Aggregating prediction across scans improves overall discrimination compared with single-scan outputs. **E.** Comparison with state-of-the-art (SOTA) models. For side-level MP prediction using B-mode images, CARN with curriculum learning (CARN + CL) achieved the highest AUC among all tested architectures ($p < 0.05$, DeLong's test). **F.** Comparison with clinical baselines. ROC curves comparing CARN against logistic-regression baselines using PPT and pain rating. With visit-level aggregation, CARN achieved AUC = 0.853, outperforming PPT (0.800) and pain-rating (0.776) models. A combined model integrating CARN + PPT + pain rating achieved AUC = 0.995, demonstrating that ultrasound-derived features provide substantial complementary diagnostic value.

## 3. Discussion

We introduce LAYER, a multimodal, XAI framework that quantitatively determines the contributions of individual and combined tissue layers to MP. This work introduces three core innovations: (1) the CARN that stabilizes training and improves classification by emphasizing informative ultrasound regions; (2) a tissue-layer-wise occlusion strategy that generates quantitative saliency scores reflecting each layer's predictive importance; and (3) an OIS that captures synergistic or redundant relationships among tissue layers. Together, these components enable anatomically grounded, quantitative explainability that extends beyond qualitative post hoc visualization methods such as Grad-CAM, Integrated Gradients, or SmoothGrad. Unlike traditional statistical or radiomic models that rely on handcrafted geometric or texture features, LAYER learns directly from volumetric ultrasound representations and provides layer-specific, interpretable, and biologically meaningful attributions of disease-relevant signal.

Our analysis revealed that non-muscle structures, particularly DFM in B-model imaging, carry meaningful predictive value for MP. This finding aligns with prior studies [22,23] implicating deep fascia and thoracolumbar fascia (TLF) in low-back pain, but also highlights previously overlooked roles for superficial fascia and dermal layers, which showed high saliency after volume normalization. We observed modality-specific patterns: DFM dominated in B-mode imaging, whereas the muscle layer was most salient in SWE. These trends reflect intrinsic imaging physics. B-mode better captures echogenic fascia and dermis, which are more superficial, while SWE more sensitively quantifies stiffness alterations within deeper muscle fibers, consistent with reports linking localized shear-wave speed elevations to trigger points [24-29]. Thus, LAYER not only localizes predictive features but also quantitatively characterizes modality-specific information flow within lumbar soft tissues.

By uniting methodological innovation with new biological insight, LAYER establishes a paradigm shift in ultrasound-based analysis of MP. It provides quantitative, anatomically interpretable measures of tissue-layer relevance, validated through faithfulness, sanity, and clinical association tests. These analyses confirm that LAYER's saliency patterns depend on learned model weights, align with physiological expectations, and are reproducible across modalities. Clinically, this framework offers a pathway toward precision rehabilitation strategies targeting the tissue layers most contributing to MP. Beyond low back pain, the methodological advances of LAYER may be applicable to other musculoskeletal and soft tissue disorders (such as in the cervical spine or non-axial muscle), extending its impact across multiple clinical domains.

We acknowledge several limitations of this study. First, the MUSCLE cohort includes a moderate number of participants (n=113). However, it contains more than 2,600 B-mode and 1,400 SWE volumes, representing, to our knowledge, the largest 3D ultrasound datasets for MP to date. Furthermore, repeated imaging across anatomic sites and multiple visits enabled capture of substantial intra-patient variability, and participant-level five-fold cross-validation prevents data leakage and supports robust generalization. The CARN architecture and the consistency of faithfulness and sanity check further reduce overfitting risk, indicating that saliency patterns reflect meaningful, tissue-specific signal rather than noise. Second, saliency magnitudes may be influenced by layer size and data quality. Spatial variations in B-mode signal-to-noise ratio (SNR) due to depth-dependent attenuation, echogenicity, or tissue composition can influence the apparent information content of a layer and bias saliency interpretation. Likewise, SWE saliency may be impacted by variability in shear-wave propagation quality and the accuracy of time-of-flight fitting. These imaging artifacts can introduce non-physiological contributions to saliency scores, and future work incorporating explicit image-quality metrics will help further mitigate these effects. To reduce this bias, we reported volume-adjusted saliency and plan to integrate explicit image-quality metrics in future work. Third, LAYER does not rely on handcrafted radiomic or geometric features. While this could be viewed as a limitation, we regard it as a strength. Traditional radiomic approaches require manual feature engineering and may miss complex, spatially distributed abnormalities. In contrast, LAYER directly learns feature representations from volumetric ultrasound and uses layer-wise occlusion analysis to generate quantitative, tissue-specific saliency scores, enabling more localized and biologically meaningful interpretation. Thus, LAYER provides greater flexibility and insight than conventional feature-based methods, with the option to integrate radiomic descriptors in future studies. Fourth, the current analysis focused on binary classification (MP or trigger-point presence). Future work will include patient-reported severity measures and longitudinal follow-up to evaluate whether LAYER can stratify disease burden, monitor therapy response, and track recovery dynamics over time. Finally, ground truth labels for the tissue layer saliency scores are not available, preventing direct validation of LAYER outputs. However, we addressed this limitation through multiple validation strategies. Faithfulness tests showed that LAYER produced the steepest insertion/deletion curves compared with existing XAI methods, confirming that identified layers drive predictions. Sanity checks with model randomization led to >1,000-fold reductions in saliency magnitudes, indicating dependence on learned rather than spurious patterns. Furthermore, directional saliency scores significantly correlated with clinical MP and trigger-point labels, and results were consistent across B-mode and SWE modalities, supporting biological plausibility. Collectively, these complementary evaluations provide rigorous indirect validation of LAYER outputs.

In summary, LAYER provides a quantitative, anatomically grounded XAI framework that dissects tissue-layer contribution in 3D ultrasound imaging of MP and trigger points. It bridges the gap between computational prediction and physiological insight, providing interpretable evidence of how individual tissues contribute to pain pathology. This framework establishes a foundation for precision ultrasound diagnostics, enabling future studies to integrate pain severity, larger cohorts, and longitudinal data to refine disease characterization, monitor therapeutic response, and advance mechanistic understanding of MP.

## 4. Methods

### 3.1 Data acquisition and curation

The dataset analyzed in this study was collected from an ongoing National Institutes of Health (NIH)-funded study, the MUSCLE study (Institutional Review Board (IRB): STUDY22090014). Participants were enrolled from an academic medical center through the Pitt+Me volunteer registry and clinic referrals from physiatry and pain medicine services.

### Participants

Eligible participants were adults with cLBP of ≥3 months' duration. Control participants reported no current cLBP. Exclusion criteria included recent lumbar surgery, acute radiculopathy, systemic rheumatologic disease, pregnancy, and other standard safety considerations. The analyzed cohort comprised 113 participants (67 female), reflecting real-world demographic distributions.

### Ultrasound acquisition

Participants were positioned prone, and scans were acquired with a 6 MHz row-column addressing (RCA) transducer (RC6gV, Verasonics, Inc., WA, USA) connected to a Vantage 256 ultrasound imaging system (Verasonics, Inc., WA, USA). The transducer was placed laterally to the midline at the L3–L4 interspace, targeting the MF and ES muscles (Fig. 1 A).

To obtain structural volumetric images using the RCA transducer with a better point spread function (PSF), three consecutive B-mode volume frames were acquired using a synthetic aperture approach with temporal compounding[30]. Co-registered SWE volumetric data were collected at the same locations. Shear waves were generated by an external vibrator excited at 150 Hz positioned superior to the imaging transducer. To achieve sufficient temporal resolution for wave propagation reconstruction, sixty B-mode volume frames were acquired using a modified orthogonal plane wave (OPW) sequence [31].

An asymmetric beamforming configuration was applied, with column-element transmission and row-element reception over 16 uniformly spaced angles (–15° to +15°), achieving a volume rate of 800 Hz. Shear-wave propagation speed was estimated at each voxel using a 3D time-of-flight based approach [32], producing volumetric shear-wave speed maps, subsequently converted to shear modulus ($\mu = \rho c^2$; $\rho = 1,000$ kg/m³). Per visit, two anatomical sites (MF and ES) were imaged on each side. At each site, three B-mode and two SWE acquisitions were obtained, yielding up to 12 B-mode and 8 SWE volumes per side per visit (and up to 24 B-mode and 16 SWE across two visits).

**Clinical labeling**

MP and trigger points were identified using standardized palpation from L5 to L3, extending from the multifidus laterally to the iliac crests. Examiners applied approximately 3–4 kgf of pressure for 5–10 seconds at predefined myotomes or taut bands. Tender points were defined as localized pain without referring or radiating pain away from the direct area of palpation. Trigger points were defined as hypersensitive nodules that reproduced the participant's characteristic pain that radiated to regions away from the area of palpation upon sustained compression.

Three board-certified physiatrists (musculoskeletal ultrasound experience 9 (ACB), 7 (RPN), and 25 (ADW) years), independently assigned side-level MP labels while blinded to elastography data. Disagreements were resolved by consensus. Each ultrasound scan was labeled according to healthy controls if no tender or trigger points were detected, or as MP present if at least one tender or trigger point was identified. MP scans were further subcategorized as tender MP (≥1 tender points, no trigger points) or trigger points (≥1 trigger points, either with/without tender points). At the clinical-visit level, labels were aggregated across both sides: a visit was considered a healthy control (visit) when no tender or trigger points were found bilaterally, and MP present (visit) when at least one tender or trigger point was detected on either side.

**3.2 Tissue-layer segmentation via GRN**

Six anatomical tissue layers were segmented from B-mode ultrasound volumes using the GRN [21] (Fig. 1 E). As illustrated in Supplementary Fig 1, GRN jointly optimizes three components: a generator $G$, a segmentation model $S$ (Segmentor), and a PatchGAN discriminator $D$ in a joint objective. The segmentation loss computed on generator-reconstructed images is backpropagated into $G$. This design encourages $G$ to produce easy-to-learn reconstructions that improve downstream segmentation accuracy. Under supervised learning, the network minimizes a composite objective:

$$\min L_{seg}\big(S\big(G(x)\big), y\big) + \lambda_1 ||G(x) - x||_1 + \lambda_{adv} L_{adv}\big(D, G(x)\big) \tag{1}$$

where $x$ is the input B-mode image, and y is its corresponding ground-truth segmentation. $L_{seg}$ represents the segmentation loss between the predicted and reference tissue masks. The L1 reconstruction term encourages anatomical fidelity between the input and reconstructed image. $L_{adv}$ denotes the adversarial loss derived from the PatchGAN discriminator, weighted by $\lambda_{adv}$. For unlabeled data, the segmentation loss $L_{seg}$ is replaced by an interpolation-consistency loss $L_{ICT}$, which enforces consistency between the prediction of interpolated images and the interpolation of their individual predictions. During inference, G and S are concatenated to produce final prediction $\hat{y} = S(G(x))$. A detailed description of GRN and its performance can be found elsewhere [21].

### 3.3 Curriculum-Adaptive Re-weighting Network (CARN)

To facilitate accurate tissue layer analysis and classification, we developed CARN, which consists of two key components: (1) **Self-paced curriculum learning:** prioritizes easy-to-learn scans during early training and gradually incorporates more difficult scans; and (2) **Adaptive Intensity Re-weighting (AIR):** applies learnable voxel-wise spatial weight maps to emphasize the most informative regions within each ultrasound scan. The 3D U-Net-based weight generator produces these maps, which are applied to the input before classification by a 3D ResNet backbone.

### 3.3.1 Self-paced curriculum learning

B-mode and SWE were first concatenated into a single multimodal input, allowing the model to exploit complementary structural and mechanical information (Fig. 4 A). For each training image $i$, an exponentially smoothed difficulty score was maintained to guide sample selection during curriculum learning:

$$L_i^{(t)} = \beta L_i^{(t-1)} + (1 - \beta) l_i^{(t)} \tag{2}$$

where $l_i^{(t)}$ is the instantaneous per-sample loss at iteration $t$, and $\beta$ is the exponential moving average (EMA) momentum. At the beginning of epoch $e$, the sampler selects the easiest $N_e$ samples as the inputs for training the classification model according to their difficulty score:

$$N_e = f_e N, f_e = f_{min} + (1 - f_{min}) \frac{e}{E-1} \tag{3}$$

where $N$ is the full training set size, $E$ is the total number of epochs, and $f_{min}$ is the initial exposure rate that determines the proportion of the dataset used in the first epoch. As training progresses, $f_e$ increases, gradually expanding the sample pool to include more challenging images. After each back-propagation step, the difficulty score $L_i^{(t)}$ is updated via Eq. (1), allowing the sampler to dynamically adjust sample selection throughout training.

### 3.3.2 Adaptive Intensity Re-weighting (AIR)

After the curriculum scheduler selects ultrasound scans, they are passed through a 3D U-Net-based weight generator to produce a voxel-wise intensity weight map (Fig. 4 A). The map is applied element-wise to the original images to emphasize anatomically and diagnostically informative regions. The re-weighted input is then fed into the classifier to produce final predictions. Formally, for an input image $I$, weights generator $G$ and classifier $C$, the prediction utilizing AIR could be expressed as:

$$\hat{y} = C\left(\left(G(I)\right) \odot I\right) \tag{4}$$

Where $\odot$ denotes element-wise multiplication. During each training iteration, the loss is backpropagated jointly through both the weight generator and the classifier, allowing the generator G to learn task-specific spatial weighting that enhances discriminative features. The same loss signal is also propagated to the curriculum scheduler to update each sample's difficulty score based on its latest classification performance.

This joint optimization of curriculum progression and adaptive reweighting ensures that the network simultaneously learns what to focus on (via AIR) and when to learn it (via self-paced scheduling), resulting in more stable convergence and anatomically meaningful feature representations.

### 3.4 Prediction aggregation across scans

Because not all scans from the same patient carry equally diagnostic value, we employ a prediction aggregation strategy to enhance overall model performance (Fig. 4 A). The classification model $M$ is trained at the scan level, where each scan is evaluated independently for the presence of MP/trigger points. During inference, multiple scans from different levels of the same patient are passed through the classifier, and their outputs are aggregated to produce a patient-level prediction. To balance performance across positive and negative cases, we adopted simple averaging as the aggregation method.

$$P = \frac{\sum_{n=1}^{N} M(I_n)}{N} \tag{5}$$

where $I_n$ denotes n-th scans from the same patient, and $N$ is the total number of images considered at the aggregation level. This approach provides stable patient-level estimates while minimizing the influence of noisy or low-quality scans.

### 3.5 Explainability with LAYER

We introduce a tissue-layer-wise occlusion strategy to quantify both single-layer and multi-layer contributions to model predictions (Fig. 2 A C).

**Single-layer saliency score**

To quantify the importance of each tissue layer, we define a *saliency score*. Let $C$ denote the classification model, $I$ the input image, and $M_i$ the mask of the $i$-th tissue layer. The saliency score for this tissue layer $i$ is computed as follows:

$$SS_i = \left| logit\big(C(I)\big) - logit\Big(C\big(I \odot (1 - M_i)\big)\Big) \right| \tag{6}$$

where $\odot$ denotes element-wise multiplication. The $logit$ function linearizes probability outputs. A higher $SS_i$ indicates greater importance of layer $i$ for MP or trigger point prediction.

**Directional saliency**

To determine the direction of this contribution, we introduce a directional saliency score for each image:

$$\Delta_i^n = logit\big(C(I)\big) - logit\Big(C\big(I \odot (1 - M_i)\big)\Big) \tag{7}$$

By aggregating the directional saliency scores across a dataset comprising $n$ images, we define the PDSS and the NDSS for tissue layer $i$ as follows:

$$PDSS_i = \frac{\sum_{n=1}^{N} max(\Delta_i^n, 0)}{N} \tag{8}$$

$$NDSS_i = -\frac{\sum_{n=1}^{N} \min(\Delta_i^n, 0)}{N} \tag{9}$$

If $PDSS > NDSS$, inclusion of layer $i$ generally shifts predictions toward MP or trigger points.

**Volume-adjusted scores**

Because large tissue layers may inflate saliency due to volume rather than information content, all scores were normalized by layer volume $\|M_i\|$ :

$$SS_i^{(voxel-adjusted)} = \frac{SS_i}{\|M_i\|} \tag{10}$$

The volume-adjusted saliency scores measure the information intensity within a specific tissue layer regardless of its volume. Similarly, the volume-adjusted positive and negative directional saliency scores are given by:

$$PDSS_i^{(voxel-adjusted)} = \frac{\sum_{n=1}^{N} \frac{max(\Delta_i^n, 0)}{\|M_i^n\|}}{N} \tag{11}$$

$$NDSS_i^{(voxel-adjusted)} = -\frac{\sum_{n=1}^{N} \frac{min(\Delta_i^n, 0)}{\|M_i^n\|}}{N} \tag{12}$$

**Inter-layer relationships**

To examine the interactions among layers, we compute the Pearson correlation coefficient between their saliency scores:

$$\rho_{i,j} = \frac{cov(SS_i, SS_j)}{\sigma_{SS_i}\sigma_{SS_j}} \tag{13}$$

where $SS_i$ and $SS_j$ represent saliency scores for layer $i$ and $j$. $cov()$ denotes covariance, and $\sigma$ is the standard deviation. We also quantify interactions between tissue layers using the OIS:

$$OIS_{i,j} = \frac{SS_{i,j} - (SS_i + SS_j)}{\max(SS_i + SS_j, \varepsilon)} \tag{14}$$

where $\varepsilon$ is a small constant to prevent division by zero. A positive $OIS_{i,j}$ indicates a synergistic effect of layers $i$ and $j$ (joint occlusion has a greater impact than expected), while a negative value suggests redundancy between layers.

### 3.6 Faithfulness, sanity, and association tests

To ensure that LAYER's layer-wise interpretations depend on the learned model representation rather than chance, we performed three complementary evaluations: a faithfulness test, a model-randomization sanity check, and an association test.

**Faithfulness test**

The faithfulness analysis quantifies how well an explainability method identifies regions that truly affect model predictions [21]. We measured how the model's output logit changes as tissue layers are cumulatively inserted or deleted according to their saliency ranks (Eq. (5)). For each image, layers were first ordered by descending saliency score. The insertion curve starts with an empty image and adds the top-k ranked layers sequentially, recording the target logit after each addition ($i_k$). The deletion curve starts from the full image and removes the top-k ranked layers in order, recording the logit after each removal ($d_k$). For these curves, we computed four quantitative metrics:

**AUC-insertion** (area under the insertion curve):

$$AUC_{ins} = \frac{1}{K} \sum_{k=1}^{K} \frac{1}{2} (i_{k-1} + i_k) \tag{15}$$

**AUC-Deletion** (area under the deletion curve):

$$AUC_{del} = \frac{1}{K} \sum_{k=1}^{K} \frac{1}{2} (d_{k-1} + d_k) \tag{16}$$

**AUC-Δ** (difference of AUCs):

$$AUC_{\Delta} = AUC_{ins} - AUC_{del} \tag{17}$$

**IROF** (increase in Region of Focus, $\varepsilon$ is a small constant):

$$IROF = \frac{max_{0 \le k \le K} i_k}{min_{0 \le k \le K} d_k + \varepsilon} \tag{18}$$

Higher AUC-insertion, AUC-Δ, IROF, and lower AUC-deletion indicate that the XAI method more faithfully identifies tissue layers that drive the model prediction.

**Model-randomization sanity check**

To verify that LAYER's explanations arise from learned weights rather than image structure, we reinitialized the pretrained CARN model with random parameters and recomputed tissue-layer saliency scores [33]. A significant drop in average saliency magnitudes between the original and randomized rankings confirms that LAYER's saliency output relies on learned weights rather than arbitrary patterns.

**Association test for directional saliency**

We next evaluated whether directional saliency scores (PDSS and NDSS) were statistically associated with MP status at the side level. Univariate logistic regression models were fitted for each tissue layer:

$$logit\big(P(MP = 1)\big) = \beta_0 + \beta_1 S_i, \qquad S_i \in (PDSS_i, NDSS_i) \tag{19}$$

For each model, we reported the coefficient ($\beta_1$), 95% confidence interval (CI), p-value, and the model's AUC. We pre-specified a directional significance criterion: $\beta_1 > 0$ with p-value $< 0.05$ for PDSS or $\beta_1 < 0$ with p-value $< 0.05$ for NDSS. Meeting this criterion indicates that larger positive (or negative) directional saliency scores from layer $i$ are associated with higher (or lower) odds of MP in the expected direction, supporting that the directional saliency scores are reliably oriented rather than driven by random variability.

### 3.7 Experiment setup

During the model training phase, CARN was trained and evaluated using k-fold cross-validation under three clinical scenarios with hierarchical prediction aggregation: (1) Clinical visit-level MP prediction (MP present vs. healthy normal controls), predictions were first aggregated at the repetition-level from scans from the same repetition, and then aggregated across the three repetitions to obtain visit-level predictions. (2) Side-level MP prediction (MP present vs. healthy normal controls): Predictions of all scans acquired on the same anatomical side (left/right) were aggregated. (3) Side-level trigger-point prediction (trigger MP vs. healthy normal controls + tender MP): Predictions of all scans acquired on the same anatomical side are aggregated.

To quantify the contributions of different ultrasound imaging modalities, CARN was trained and tested using B-mode images alone, SWE alone, and concatenated "B-mode and SWE". After training, a randomly selected fold-specific CARN checkpoints were used in the interpretation phase to perform tissue-layer-wise occlusion, including single-layer analysis, multi-layer analysis, and extraction of saliency maps on the validation dataset.

CARN was benchmarked against four state-of-the-art methods on our study cohort, including 3D-CNN, ViT [34], Swin Transformer [35], and ResNet. Metrics include precision, recall, F1-score, accuracy, and AUC. To quantify LAYER's interpretability, we performed a comprehensive XAI faithfulness test against Grad-CAM[18], IG[36], and SmoothGrad[37] using faithfulness metrics (AUC-Insertion, AUC-Deletion, AUC-Δ and IROF).

In all experiments, the weights generator was a 3D U-Net model with feature dimensions of 32, 64, 128, and 256, and the classifier was a 3D ResNet-50 model. Models were implemented in PyTorch 2.3.1 and optimized with the Adam optimizer (learning rate = 0.0001). To stabilize training and prioritize easier samples, we employed the EasyFirstSampler, which selects samples with lower current difficulty (estimated via an exponential moving average, EMA). The initial exposure rate was set to 0.2, and EMA momentum to 0.9. Input ultrasound images were zero-padded. All computations were performed on an NVIDIA GTX3090 GPU.

**Code availability.** The deep-learning models were developed using standard libraries and scripts available in Pytorch. Code is available at https://github.com/Francisdadada/LAYER.

**Data availability:** The dataset for this study is protected patient information. Some data may be available for research purposes from the corresponding author upon reasonable request.

**Acknowledgements:**

This research was supported in part by the University of Pittsburgh Center for Research Computing and Data, RRID:SCR_022735, through the resources provided. Specifically, this work used the H2P cluster, which is supported by NSF award number OAC-2117681. This work used the Open Storage Network (OSN) through allocation MED230013 from the Advanced Cyberinfrastructure Coordination Ecosystem: Services & Support (ACCESS) program, which is supported by National Science Foundation grants #2138259, #2138286, #2138307, #2137603, and #2138296. This work is supported in part by research grants from the National Institutes of Health (NIH) (R61AT012282).

**Author contributions:**

Z. Z., J. P. initiated and designed the research. Z. Z. executed the research. Z. S., M. S., J. C., A. B., R. N., E. L., K. L., A. W., J. P. acquired the data. K. K., A. W., J. P. supervised the data collection. Z. Z. analyzed and interpreted the data. Z. Z. developed algorithms and software tools for the experiment. Z. Z., A. P., T. Y., G. K., H. L., J. W., X. M., Z. S., A. W., J. P. wrote the manuscript.

**Competing interests:**

Authors declare no competing interests.